\documentstyle[multicol,prl,aps,epsf]{revtex}
\begin{document}
\draft
\title{Timesaving Double-Grid Method for Real-Space Electronic-Structure Calculations}
\author{Tomoya Ono and Kikuji Hirose}
\address{Department of Precision Science and Technology, Osaka University, Suita, Osaka 565-0871, Japan}
\date{\today}
\maketitle
\begin{abstract}
We present a simple and efficient technique in {\it ab initio} electronic-structure calculation utilizing real-space double-grid with a high density of grid points in the vicinity of nuclei. This technique promises to greatly reduce the overhead for performing the integrals that involves non-local parts of pseudopotentials, with keeping a high degree of accuracy. Our procedure gives rise to no Pulay forces, unlike other real-space methods using adaptive coordinates. Moreover, we demonstrate the potential power of the method by calculating several properties of atoms and molecules.
\end{abstract}

\begin{multicols}{2}
\narrowtext

So far, a number of methods for {\it ab initio} electronic-structure calculations entirly in real space \onlinecite{rs1,rs2,rs3,rs4,rs5,rs6,gygi,rs7} have been proposed. They have some advantages compared with the usual plane-wave approach. The first one is that boundary conditions are not constrained to be periodic, e.g., nonperiodic boundary conditions for molecules and a combination of periodic and nonperiodic boundary conditions for surfaces. Even more important is that a technique utilizing real-space double-grid is available where many more grid points are put in the vicinity of nuclei, so that the integrals involving rapidly varying pseudopotentials inside the core regions of atoms can be calculated with a high degree of accuracy. This kind of integration over numerous sampling points, however, requires a large sum of computational effort. In this Letter, we present a quite simple and efficient double-grid technique which yields a drastic reduction of the computational cost without a loss of accuracy in the framework of the real-space finite-difference method. This technique can also be applied to the plane-wave approach, if the integration over the core region is implemented in real space.

The double-grid employed here consists of two sorts of uniform and equi-interval grid points, i.e., coarse- and dense-ones, depicted in Fig.~\ref{fig:1} by the marks ``$\times$'' and ``$\bullet$'', respectively. The dense-grid region enclosed by the circle is the core region of an atom that is taken to be large enough to contain the cutoff region of non-local parts of pseudopotentials. Throughout this paper we postulate that wave-functions are defined and updated {\it only on coarse-grid points}, while pseudopotentials are strictly given {\it on all dense-grid points} in an analytically or numerically exact manner.

Let us consider inner products between wave-functions $\psi(x)$ and non-local parts of pseudopotentials $v(x)$ (see Fig.~\ref{fig:2}). For simplicity, the illustration is limited to the one-dimensional case hereafter. The values of wave-functions on coarse-grid points ($\circ$) are stored in computer memory, and the values on dense-grid points ($\bullet$) are evaluated by interpolation from them. The well-known values of pseudopotentials both on coarse- and dense-grid points ($\circ$) are also shown schematically. Then, from Fig.~\ref{fig:2}(a) one can see that only the values on coarse-grid points are so inadequate that the inner products can not be accurately calculated; the errors are mainly due to the rapidly varying behavior of pseudopotentials. On the other hand, Fig.~\ref{fig:2}(b) indicates that the inner products can be evaluated to great accuracy, if the number of dense-grid points is taken to be sufficiently large and also if the values of wave-functions on dense-grid points are properly interpolated from those on coarse-grid points \cite{comment1}.

There are several interpolation methods for wave-functions, among which the simplest is linear interpolation. In this case, the values of wave-functions $\psi_j \equiv \psi(x_j)$ on dense-grid points $x_j$ are interpolated from the values $\Psi_J \equiv \psi(X_J)$ on coarse-grid points $X_J$ as
\begin{equation}
\label{eqn:linear}
\psi_j = \frac{h - (x_j-X_{J})}h \Psi_{J} + \frac{h - (X_{J+1} -x_j)}h \Psi_{J+1}  ,
\end{equation}
where $h$ is the grid spacing of coarse-grid points. The inner product is assumed to be accurately approximated by the discrete sum over dense-grid points, i.e.,
\begin{equation}
\label{eqn:inner}
\int^{d/2+R_I}_{-d/2+R_I} v^I(x) \psi(x) dx \approx \sum_{j=-nN_{core}}^{nN_{core}} v^I_j\psi_j h_{dens},
\end{equation}
where $d$ is the ``diameter'' of the core region, $R_I$ is the atomic position, $v^I_j \equiv v^I(x_j) = v(x_j -R_I)$, $2N_{core}+1$ ($2nN_{core}+1$) is the number of coarse- (dense-) grid points in the core region, $h_{dens}$ is the grid spacing of dense-grid points, and $n=h/h_{dens}$, i.e., $n-1$ is the number of dense-grid points existing between adjacent coarse-grid ones. Now, substituting Eq.(\ref{eqn:linear}) in the r.h.s. of Eq.(\ref{eqn:inner}), we have
\begin{equation}
\label{eqn:kekka}
\int^{d/2+R_I}_{-d/2+R_I} v^I(x) \psi(x) dx \approx \sum_{J=-N_{core}}^{N_{core}} w^I_J\Psi_J h,
\end{equation}
where
\begin{equation}
w^I_J=\sum_{s=-n}^n \frac{h - |x_{nJ+s} - X_J|}{nh}v^I_{nJ+s}.
\end{equation}
As shown in Eq.(\ref{eqn:kekka}), the r.h.s. of the inner product (\ref{eqn:inner}) has been replaced with {\it the summation over coarse-grid points inside the core region}, which produces only a modest overhead in the computational cost. It should be remarked that the weight factors $w^I_J$ arising from the interpolation are independent of the wave-functions, but dependent only on the well-known values of pseudopotentials on dense-grid points. Thus, if once computing the factors $w^I_J$ every molecular-dynamics time-step, we do not have to recalculate them throughout self-consistent iteration-steps. The extension of the above procedure to the cases of higher-order interpolations is straightforward.

Fourier interpolation is of great interest, since the term representing the position of the atom can be factorized in the expression of $w^I_J$ as the structural phase factor $\exp(-ik\frac{2\pi}{d}R_I)$. Indeed, the interpolated values of wave-functions $\psi_j$ on dense-grid points $x_j$, i.e.,
\begin{mathletters}
\label{eqn:fourier}
\begin{equation}
\label{eqn:fourier:1}
\psi_j = \sum_{k=-N_{core}}^{N_{core}} \tilde{\Psi}_k e^{ik\frac{2\pi}{d}x_j},
\end{equation}
with
\begin{equation}
\label{eqn:fourier:2}
\tilde{\Psi}_k = \frac{1}{d} \sum_{J=-N_{core}}^{N_{core}} \Psi_J e^{-ik\frac{2\pi}{d}X_J} h,
\end{equation}
\end{mathletters}
are substituted into the r.h.s. of Eq.(\ref{eqn:inner}) to give Eq.(\ref{eqn:kekka}), where the weight factors in this case are
\begin{mathletters}
\label{eqn:ft}
\begin{equation}
\label{eqn:ft:1}
w^I_J = \sum_{k=-N_{core}}^{N_{core}} \tilde{w}_k e^{ik\frac{2\pi}{d}X_J} e^{-ik\frac{2\pi}{d}R_I},
\end{equation}
with
\begin{equation}
\label{eqn:ft:2}
\tilde{w}_k = \frac{1}{d} \sum_{j=-nN_{core}}^{nN_{core}} v_j e^{-ik\frac{2\pi}{d}x_j} h_{dens}.
\end{equation}
\end{mathletters}
Here $v_j \equiv v(x_j)$. An advantage to be stressed is that the calculation of making the table of $\tilde{w}_k$ in Eq.(\ref{eqn:ft:2}) for each atom species, which requires a time-consuming computational effort because of the summation over dense-grid points, has only to be carried out once at the early stage of the entire job. At that time, the usage of the fast Fourier transforms (FFT) will considerably reduce an amount of computational operation.

We now examine the performance of our method through calculations of several properties of atoms and molecules. Hereafter, we obey the nine-points finite difference formula (i.e., the formula with N=4) \cite{rs3} for the derivatives arising from the kinetic-energy operator, imposing a nonperiodic boundary condition of vanishing wave-functions. The dense-grid spacing is fixed at $h_{dens}=h/9$. The electronic charge-density, the Hartree potential, and the exchange-correlation potential are assumed to be described only on coarse-grid points. Exchange-correlation effects are treated using the local-spin-density approximation \cite{lda} of the density-functional theory. The norm-conserving pseudopotential of Bachelet, Hamann, and Schl\"uter (BHS) \cite{hsc} is employed in a separable non-local form \cite{sep}.

The convergence of the total energy for the hydrogen atom as a function of the coarse-grid cutoff energy is presented in Fig.~\ref{fig:3}. According to Ref.\cite{gygi}, we defined an equivalent energy cutoff $E_c^{coars}$ [$\equiv (\pi/h)^2$ Ry] to be equal to that of the plane-wave method which uses a FFT grid with the same spacing as the present calculation. The position where the atom is located is at the center between adjacent grid points. The hydrogen atom is one of the most difficult atoms to treat, owing to the rapid oscillation of its s-state non-local pseudopotential, and consequently the total energy as a function of the cutoff energy calculated without any interpolation sharply oscillates. On the other hand, our prescription with interpolations drastically improves the results; the obtained total energies converge rapidly and monotonously as the cutoff energy increases.

Fig.~\ref{fig:4} shows the total-energy variation as a function of the displacement of the oxygen atom relative to coarse-grid points along a coordinate axis. The coarse-grid spacing is taken to be $h=0.21$ a.u. The first-row elements are difficult atoms to deal with, because their pseudopotentials are rapidly varying in the cutoff region. In treating these elements in the context with a real-space approach, it is necessary to consider {\it the dependence of the energy on the position of the atom}. As seen in Fig.~\ref{fig:4}, the energy variation in our scheme with cubic interpolation is $\sim$0.04\% of the total energy, which is negligibly small.

We next apply our method to the calculation of the equilibrium bond length of the CO$_2$ molecule in order to examine the performance of our method for molecules. The total energies as a function of the C-O distance are shown in Fig.~\ref{fig:5}. We employ the cubic interpolation formula of the present method and take the coarse-grid spacing $h$ to be 0.21 a.u. The results without any interpolation have many ``humps'', and the distances between adjacent humps on the respective dotted curves are close to the grid spacing $h$, which confirms that the oscillation depends on the relative position of the atom with respect to coarse-grid points. To circumvent this problem, Gygi {\it et al.} \cite{gygi} proposed a real-space grid in adaptive curvilinear coordinates. However, owing to the use of these distorted coordinates, their scheme needs {\it Pulay forces} in molecular-dynamics simulations; it would be computationally very demanding. On the contrary, our approach requires only the Hellmann-Feynman forces, which significantly reduces the computational cost concerning the calculation of forces. As shown in Fig.~\ref{fig:5}, the total energies computed with our method do not make absurd dependence on the position of the atom. The bond length determined by the minimum of the total energy is in excellent agreement with the experimental data 2.19 a.u. These results make it clear that our method solves this problem and that it is very efficient and applicable \cite{comment2}.

Finally, the calculated properties of the other molecules are given in Table \ref{tbl:1}. The coarse-grid resolution $h$ is 0.24 a.u. for N$_2$ and 0.21 a.u. for the other molecules, and the cubic interpolation formula is used. Our results are in good agreement with those of experiments and other theories.

In summary, we have presented a method for performing the finite-difference electronic-structure calculations using real-space double-grid. Our method has the following desirable features: (i)The inner products between wave-functions and pseudopotentials are evaluated with the same accuracy as in the case of dense-grid points, in spite of the calculation with respect to coarse-grid points. (ii)The computational effort is modest, thanks to the integration over coarse-grid points. (iii)The double-grid acts as {\it stabilizer} in the calculation of the total energy, i.e., it suppresses the spurious oscillation of the energy for the displacement of the atom relative to coarse-grid points. (iv)Unlike other real-space methods using adaptive coordinates, our procedure gives rise to no Pulay forces, and so a substantial increase in the computational cost does not occur. From what has been mentioned above, it seems reasonable to conclude that our method is suitable, because of its simplicity and efficiency, for large-scale molecular-dynamics simulations incorporating the norm-conserving pseudopotentials. Work is in progress to apply the method to large-scale molecular-dynamics simulations.

This work was partially supported by a Grant-in-Aid for COE Research. The numerical calculation was carried out by the computer facilities at the Institute for Solid State Physics at the University of Tokyo. Thanks are due to Masako Inagaki and Kouji Inagaki for reading the entire text in its original form.

\begin{figure}
\caption{Double-grid adopted in the text. The marks ``$\times$'' and ``$\bullet$'' correspond to coarse- and dense-grid points, respectively. The circle shows the core region of an atom which is taken to be large enough to contain the cutoff region of non-local pseudopotentials.}
\label{fig:1}
\end{figure}

\begin{figure}
\caption{Functions on coarse- and dense-grid points in the one-dimensional case. $X_J$ ($x_j$) represents a coarse- (dense-) grid point with $j=nJ+s (0 \leq s < n)$, and so $X_J=x_{nj}$. (a) The inner product between wave-function $\psi(x)$ and non-local part of pseudopotential $v(x)$ evaluated over coarse-grid points. (b) The inner product evaluated over dense-grid points.}
\label{fig:2}
\end{figure}

\begin{figure}
\caption{Convergence of the total energy for the hydrogen atom as a function of the coarse-grid cutoff energy $E_c^{coars}$. The atom is located at the center between adjacent coarse-grid points.}
\label{fig:3}
\end{figure}

\begin{figure}
\caption{Variation of the total energy as a function of the displacement of the oxygen atom relative to coarse-grid points along a coordinate axis. The coarse-grid spacing $h$ is 0.21 a.u. The result with linear interpolation is coincident with that of the cubic case within the relative error of 0.5\%.}
\label{fig:4}
\end{figure}

\begin{figure}
\caption{Total energy of the CO$_2$ molecule in the symmetric linear conformation and location of the carbon atom. (a)The carbon atom is located at the center between adjacent coarse-grid planes in each direction. The marks ``$\times$'' represent coarse-grid points. (b)The carbon atom is located on the coarse-grid plane which is at right angles to the bonding axis and at the center between adjacent coarse-grid planes in the other directions. (c)Total energy of the CO$_2$ molecule as a function of the C-O distance. The coarse-grid spacing $h$ of 0.21 a.u. is the same size as the corresponding Euclidean grid spacing in the adaptive coordinate approach \protect\cite{gygi}. The calculated points are fit to spline functions as a guide to the eye.}
\label{fig:5}
\end{figure}

\narrowtext
\begin{table}
\caption{Properties of diatomic molecules. The experimental data are from \protect\cite{exp}. The theoretical results are from our present calculation and from other methods using similar forms for the local-spin-density approximation.}
\begin{tabular}{lccc}
&N$_2$&O$_2$&CO \\ \hline
Bond length (a.u.)&&& \\
\hspace{5mm}Experiment  &2.07&2.28&2.13 \\
\hspace{5mm}This work   &2.07&2.28&2.13 \\
\hspace{5mm}Other theory&2.07\tablenotemark[2]&2.27\tablenotemark[2]&2.13\tablenotemark[1] \\
Vibrational mode (cm$^{-1}$)&&& \\
\hspace{5mm}Experiment  &2358&1580&2169 \\
\hspace{5mm}This work   &2390&1500&2200 \\
\hspace{5mm}Other theory&2380\tablenotemark[2]&1620\tablenotemark[2]&2151\tablenotemark[1] \\
Cohesive energy (eV)&&& \\
\hspace{5mm}Experiment  &9.76&5.12&11.09 \\
\hspace{5mm}This work   &11.20&7.57&12.80 \\
\hspace{5mm}Other theory&11.6\tablenotemark[2]&7.6\tablenotemark[2]&12.8\tablenotemark[2] \\
\end{tabular}
\tablenotetext[1]{From Ref.\cite{gygi}.}
\tablenotetext[2]{From Ref.\cite{becke}.}
\label{tbl:1}
\end{table}
\end{multicols}
\end{document}